\documentclass[aps,prd,preprintnumbers,showpacs,twocolumn,groupedaddress,nofootinbib]{revtex4}
\usepackage{graphicx}
\usepackage{latexsym}
\def\beq{\begin{equation}}
\def\eeq{\end{equation}}
\def\bey{\begin{eqnarray}}
\def\eey{\end{eqnarray}}

\def\lsim{\mathrel{\raise.3ex\hbox{$<$\kern-.75em\lower1ex\hbox{$\sim$}}}}
\def\gsim{\mathrel{\raise.3ex\hbox{$>$\kern-.75em\lower1ex\hbox{$\sim$}}}}

\begin{document}

\author{Dan Hooper$^{1,2,3}$, Farinaldo S.~Queiroz$^{1,4}$, and Nickolay Y.~Gnedin$^{1,2,3}$}
\title{Non-Thermal Dark Matter Mimicking An Additional Neutrino Species In The Early Universe}  
\affiliation{$^1$Center for Particle Astrophysics, Fermi National Accelerator Laboratory, Batavia, IL 60510, USA}
\affiliation{$^2$Department of Astronomy and Astrophysics, University of Chicago, Chicago, IL 60637, USA}
\affiliation{$^3$Kavli Institute for Cosmological Physics, University of Chicago, Chicago, IL 60637 USA}
\affiliation{$^4$Departamento de F\'{i}sica, Universidade Federal da Para\'{i}ba, Caira Postal 5008, 58051-970, Jo\~{a}o Pessoa, PB, Brasil}

\date{\today}

\begin{abstract}

The South Pole Telescope (SPT), Atacama Cosmology Telescope (ACT), and Wilkinson Microwave Anisotropy Probe (WMAP) have each reported measurements of the cosmic microwave background's (CMB) angular power spectrum which favor the existence of roughly one additional neutrino species, in addition to the three contained in the standard model of particle physics. Neutrinos influence the CMB by contributing to the radiation density, which alters the expansion rate of the universe during the epoch leading up to recombination. In this paper, we consider an alternative possibility that the excess kinetic energy implied by these measurements was possessed by dark matter particles that were produced through a non-thermal mechanism, such as late-time decays. In particular, we find that if a small fraction ($\lsim 1\%$) of the dark matter in the universe today were produced through the decays of a heavy and relatively long-lived state, the expansion history of the universe can be indistinguishable from that predicted in the standard cosmological model with an additional neutrino. Furthermore, if these decays take place after the completion of big bang nucleosynthesis, this scenario can avoid tension with the value of three neutrino species preferred by measurements of the light element abundances.

\end{abstract}

\pacs{95.35.+d, 98.70.Vc; FERMILAB-PUB-11-631-A}
\maketitle

\section{Introduction}

Measurements of the temperature anisotropy of the cosmic microwave background (CMB) have revealed less power at small angular scales than is predicted in the standard cosmological model. Such a damping of small-scale power is generally interpreted as a measurement of the number of effective neutrino species, $N_{\rm Eff}^{\nu}$. Whereas the combination of the standard cosmological model and the standard model of particle physics predict a value of $N^{\nu}_{\rm Eff}$=3.04~\cite{mangano} (corresponding to the three known species of neutrinos), the WMAP collaboration has reported a measurement of $N_{\rm Eff}^{\nu}=4.34^{+0.86}_{-0.88}$, including information from measurements of the Hubble constant, and baryon acoustic oscillations~\cite{WMAP}. Similarly, the Atacama Cosmology Telescope (ACT) reports a value of $N_{\rm Eff}^{\nu}=4.6 \pm 0.8$~\cite{ACT}, and the South Pole Telescope arrives at $N_{\rm Eff}^{\nu}=3.86 \pm 0.42$~\cite{SPT}. And although none of these measurements individually deviates from the standard value by more than about two standard deviations, they collectively rule out $N_{\rm Eff}^{\nu}=3.04$ at the approximately 99\% confidence level, and instead prefer roughly one extra effective neutrinos species, $\Delta N_{\rm Eff}^{\nu}\sim0.5$-$1.6$. With the first cosmology results from the Planck satellite anticipated in early 2013, the measurement of this quantity is expected to become considerably more precise in the relatively near future.

The existence of any additional neutrino species impacts the observed anisotropies of the CMB by altering the expansion history of the universe in the epoch prior to recombination~\cite{knox}. Extra neutrinos, however, are not the only type of new particle physics which could impact the radiation density and expansion history of our universe at early times~\cite{similar}. In particular, one could consider massive particles which are produced non-thermally, such as through the decays of much heavier states. The particles produced in such decays can be highly relativistic, and thus behave as radiation until their kinetic energy is lost through cosmological redshifting. Such behavior can be found in superWIMP scenarios~\cite{Feng:2003xh}, for example. More generally speaking, we can consider any heavy state which decays to a lighter, stable state, which makes up all or some of our Universe's dark matter~\cite{decays}.

An interesting consequence of this late-decaying particle scenario is that if the lifetime of the decaying particle is longer than $\sim$$10^{3}$ seconds, the expansion history of the universe during the era of Big Bang Nucleosynthesis (BBN) will be unchanged from the standard (three neutrino) case. As measurements of the light element abundances do not provide support for the existence of additional neutrinos, and can be used to conservatively exclude $\Delta N_{\rm Eff}^{\nu}>1$ at the 95\% confidence level~\cite{serpico} (see also, however, Ref.~\cite{hamann}), we consider it well motivated to consider non-neutrino explanations for the observed lack of small scale power in the CMB.

In this paper, we consider scenarios in which a small fraction of the universe's dark matter is produced through the decays of much heavier particles. The kinetic energy of this fraction of the dark matter can alter the expansion history of the universe in a way very similar to an additional light neutrino species, potentially providing an alternative explanation for the lack of small-scale power in the temperature anisotropies of the CMB, as observed by WMAP, ACT and SPT.


\section{The Impact of Non-Thermal Dark Matter on the Expansion History of the Early Universe}

In the standard thermal history of a weakly interacting massive particle (WIMP) species, freeze-out occurs at a temperature of $T_{\rm FO} \sim m_{\rm DM}/20$~\cite{kolbturner}. As a result, the dark matter particles are only mildly relativistic at thermal freeze-out, and are highly non-relativistic by the time of matter-radiation equality. Such cold particles thus do not contribute to the radiation density of the universe at this time or to the quantity $\Delta N_{\rm Eff}^{\nu}$. This need not be the case, however, if the dark matter (or some fraction of the dark matter) were produced through a non-thermal mechanism, such as decays of much heavier, and relatively long-lived states.

At the time of matter-radiation equality, the ratio of the energy density in the three (massless) neutrinos to that in cold dark matter is given by:
\begin{equation}
\frac{\rho_{\nu}}{\rho_{\rm CDM}} = 0.690 \,\, \frac{\Omega_{\rm CMB}}{\Omega_{\rm CDM}}\, \frac{N_{\nu}}{3}\, \frac{1}{a_{\rm EQ}} \approx 0.49,
\end{equation}
where $\Omega_{\rm CMB}\approx 0.0000484$, $\Omega_{\rm CDM}\approx 0.227$, $N_{\nu}=3$ is the number of neutrino species, and $a_{\rm EQ}=3.00 \times 10^{-4}$ is the scale factor at matter-radiation equality.  The energy density in one neutrino species at equality is thus approximately equal to 16\% of the density in cold dark matter. Thus if the dark matter, instead of being entirely cold, had a kinetic energy equivalent to $\gamma_{X} \approx 1.16$ at the time of equality (and had been cooling through hubble expansion well prior to the time of equality) then it would lead to the same expansion history as predicted in the standard cold dark matter case with four (rather than three) species of neutrinos.

To explore how the dark matter may have possessed such kinetic energy at this time, consider, for example, a heavy and relatively long-lived state, $X'$, which decays, among other particles, to the particle which constitutes the dark matter of our universe, $X$. If, for concreteness, we consider a two-body decay, such as $X' \rightarrow X + \gamma$, or $X' \rightarrow X + \nu$, the Lorentz factor of the $X$ particles as a function of scale factor is given by:
\begin{equation}
\gamma_{X}(a) \approx 1 + \bigg(\frac{a(\tau)}{a}\bigg) \, \bigg(\frac{m_{X'}}{2 m_X}+\frac{m_{X}}{2 m_{X'}}-1\bigg),
\end{equation}
which makes the approximation that all of the decays happened at a time $\tau$.
During the era of radiation domination, this can be written as
\begin{eqnarray}
\gamma_{X}(t) &\approx& 1 + \bigg(\frac{\tau}{t}\bigg)^{1/2} \, \bigg(\frac{m_{X'}}{2 m_X}+\frac{m_{X}}{2 m_{X'}}-1\bigg).
\end{eqnarray}
\newpage
\noindent
When this is evaluated at the time of matter-radiation equality, we find
\begin{eqnarray}
\gamma_{X}(t_{\rm EQ})&\approx& \\
1 &+& 7.8\times 10^{-4}  \, \bigg(\frac{\tau}{10^6\,\rm{s}}\bigg)^{1/2} \,  \bigg(\frac{m_{X'}}{2 m_X}+\frac{m_{X}}{2 m_{X'}}-1\bigg).\nonumber
\end{eqnarray}
We can then relate the lifetime and masses of these particles to the equivalent number of effective neutrino species it would mimic on the expansion history:
\begin{eqnarray}
\Delta N^{\nu}_{\rm Eff} \approx 4.8\times 10^{-3} \,  \bigg(\frac{\tau}{10^6\, \rm{s}}\bigg)^{1/2} \, \bigg(\frac{m_{X'}}{m_X}+\frac{m_{X}}{m_{X'}}-2\bigg) \, f, \nonumber \\
\label{neff}
\end{eqnarray}
where $f$ is the fraction of the dark matter particles that originate from this non-thermal origin (we assume that the remainder of this fraction is non-relativistic, as expected for dark matter with a thermal origin, for example). Here we have assumed that the decay products other than the $X$ particle also contribute to the radiation density of the universe.

\begin{figure}[!t]
\centering
{\includegraphics[angle=0.0,width=3.4in]{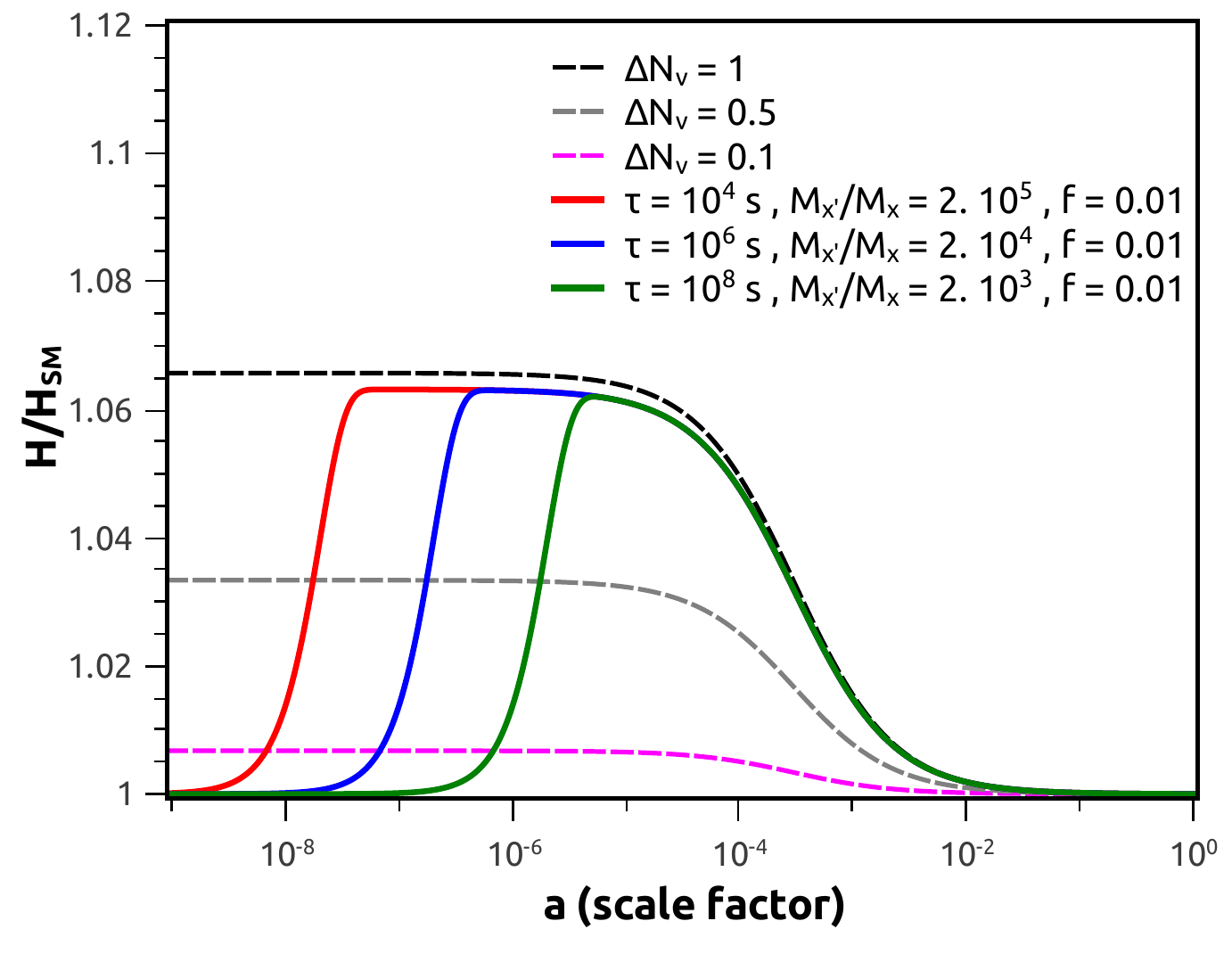}}
{\includegraphics[angle=0.0,width=3.4in]{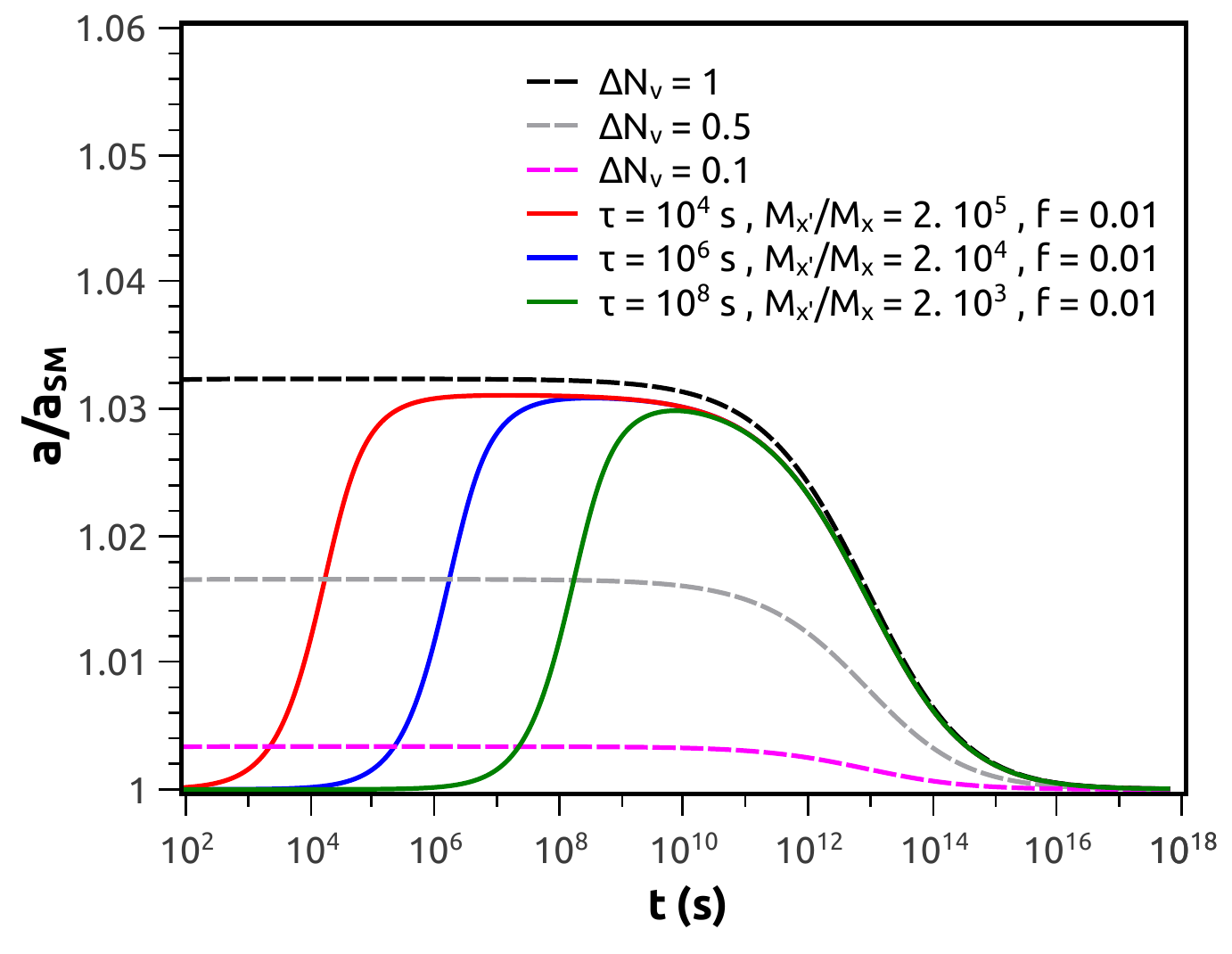}}
\caption{The effect on the universe's expansion history of additional neutrino species, and of late-time decays. In the upper frame we show the fractional change of the Hubble constant as a function of scale factor. In the lower frame, we show the fractional change in the scale factor as a function of time. In each case, we show results for three late-time decay scenarios (solid curves). We compare these results to that predicted from additional neutrino species (dashed curves) and find that these three scenarios can each effectively mimic the presence of approximately one additional light neutrino species in the early universe. Note that in each case shown, the relativistic decay products, $X$, only make up a fraction $f=0.01$ of the total dark matter density.}
\label{fig1}
\end{figure}

In Fig.~\ref{fig1}, we plot the fractional change to the expansion history of the universe for three late-time decay scenarios (solid curves). We compare these results to that predicted from additional neutrino species (dashed curves) and find that these three scenarios can each effectively mimic the presence of approximately one additional light neutrino species in the early universe.

\section{Constraints}

\subsection{Large Scale Structure}

If much of the universe's dark matter contains significant kinetic energy at the time of matter-radiation equality, the formation of large scale structure will be suppressed. In this subsection, we discuss the impact of dark matter produced in late-decays on large scale structure, and use these result to place constraints on the scenario being discussed here.


Constraints from large-scale structure evolution come from the fact that hot dark matter does not cluster below its free-streaming length, hence on sufficiently small scales the growth of perturbations is slowed down compared to a pure cold dark matter case. The free-streaming length of non-thermally generated dark matter particles (assumed to be highly relativistic after the decay, but non-relativistic by matter-radiation equality) is given by~\cite{Cembranos:2005us}:
\begin{eqnarray}
\lambda_{\rm FS} &\approx& 1.0 \, {\rm Mpc} \,  \bigg(\frac{\tau}{10^6\, \rm{sec}}\bigg)^{1/2} \, \bigg(\frac{m_{X'}}{2 m_X}-\frac{m_X}{2m_{X'}}\bigg) \\
&\times& \left\{1+0.14 \ln\bigg[ \bigg(\frac{10^6\, \rm{sec}}{\tau}\bigg)^{1/2} \, \bigg(\frac{2 m_{X'} m_{X}}{m^2_{X'}-m^2_X}\bigg)\bigg] \right\}. \nonumber
\end{eqnarray}
For parameters which lead to the impact of one additional neutrino species on the expansion history (and for $f=1$), the free-streaming length is $\lambda \sim 50$ Mpc, in considerable excess of constraints from the lyman-alpha forest ($\lambda_{\rm FS} \lsim 0.3$ to 0.07 Mpc~\cite{Viel:2010bn}). In light of this, we are forced to consider scenarios in which most of the dark matter is cold, and thus able to generate the observed large-scale structure, while a small fraction is very hot ($f \ll 1$), and potentially able to impact the expansion history of the early universe.



In linear regime, the evolution of the power spectrum of matter fluctuations has been studied by Ma (1996)~\cite{Ma:1996za}. Specifically, she finds that at scales below the free-streaming length the linear matter fluctuation evolves during the matter-dominated epoch as
\begin{equation}
  \delta \propto a^{\alpha_\infty},
\end{equation}
where $\alpha_\infty$ is given by Eq.~(3) of Ref.~\cite{Ma:1996za},
\begin{equation}
  \alpha_\infty = \frac{5}{4}\sqrt{1-\frac{24}{25}f} - \frac{1}{4} \approx
  1 - \frac{3}{5}f
\end{equation}
for sufficiently small $f$. Hence, the suppression of the small-scale power relative to the pure cold dark matter ($f=0$) case is given by
\begin{equation}
  g \equiv \frac{\delta_f}{\delta_{f=0}} = \left(\frac{a_{\rm EQ}}{a}\right)^{-\frac{3}{5}f} \approx \exp(-4.9\,f).
\end{equation}
While this expression is formally valid only in the matter-dominated regime, its correction for the cosmological-constant-dominated regime are sufficiently small as to be not important for us here.

The detailed computation of the evolution of matter clustering in our cosmological model is well beyond the scope of this paper. Thus, in order to approximately account for the large-scale structure constrains, we adopt a limit of $g>0.95$, which is broadly consistent with both the measurements of the amplitude of matter clustering (so called $\sigma_8$) from the combination of CMB constraints from WMAP 7-year data and large-scale clustering of galaxies~\cite{WMAP} and with the measured clustering power in the Lyman-alpha forest (see, for example, Ref.~\cite{Viel:2010bn}). The constraint $g>0.95$ translates into $f<0.01$ constraint on the fraction of hot dark matter.

%
%

\begin{figure}[!t]
\centering
{\includegraphics[angle=0.0,width=3.4in]{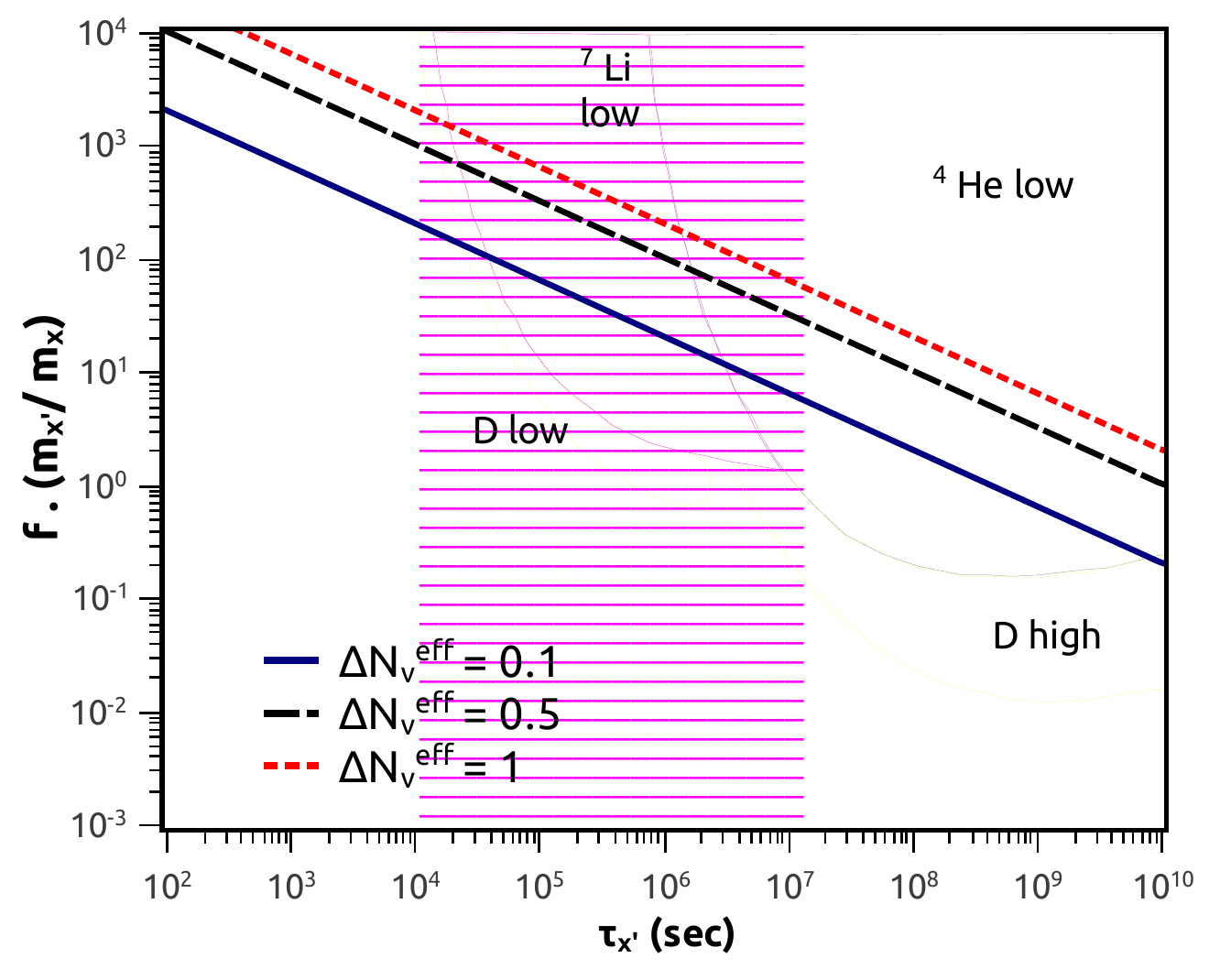}}
\caption{Constraints on the late-time decay scenario discussed in this paper from measurements of the light element abundances~\cite{bbn}. Shown for comparison are the contours which correspond to the parameter space which can mimic 1, 0.5, or 0.1 additional effective neutrino species. These constraints apply specifically to decays of the form $X' \rightarrow X + \gamma$, and can be evaded in other cases ($X' \rightarrow X + \nu$, for example). Results here are shown assuming $m_{X'} \gg m_{X}$.}
\label{fig2}
\end{figure}


\subsection{Big Bang Nucleosynthesis}

Late-time decays can potentially have a devastating impact on the successful predictions of the light element abundances. The energy of photons produced in such decays, for example, is quickly transferred through scattering with much lower energy background photons into electron-positron pairs. The resulting electromagnetic cascades can break up light nuclei, significantly altering their relative abundances.

The total electromagnetic energy released in decays of the form $X' \rightarrow X + \gamma$ is given by:
\begin{equation}
\zeta_{\rm EM} = \epsilon_{\gamma} \, Y_{\gamma},
\end{equation}
where $\epsilon_{\gamma}=(m_{X'}/2) - (m^2_X/2m_{X'})$ is the initial energy of the photon decay product and $Y_{\gamma}$ is the ratio of the number density of photon decay products to the number density of background photons. For the $X'\rightarrow X+ \gamma$ scenario being considered here, the energy release in photon decay products is given by:
\begin{equation}
\zeta_{\rm EM} \approx 1.5 \times 10^{-9}\, {\rm GeV} \, \bigg(\frac{m_{X'}}{m_X}-\frac{m_{X}}{m_{X'}}\bigg) \, f,
\end{equation}
which is $\zeta_{\rm EM} \sim 3\times 10^{-6}$ to $3\times 10^{-8}$ GeV for the parameter choices shown in Fig.~\ref{fig1}. As can be seen in Fig.~\ref{fig2}, such values of $\zeta_{\rm EM}$ can safely avoid unacceptably altering the primordial light element abundance~\cite{bbn} for $\tau \lsim 10^4$ seconds, but are in considerable conflict for longer lifetimes.  

To otherwise evade these constraints, we can instead consider decays which do not include a photon in the final state, such as $X' \rightarrow X + \nu$, or $X' \rightarrow X + X$, for example. In cases such as these, much longer lifetimes are acceptable. 

So far in this subsection, we have only discussed way in which the decay products in our scenario can effect the surviving light element abundances. In addition to this effect, the presence of any additional radiation (whether in the form of neutrinos or relativistic decay products) during the process of Big Bang Nucleosynthesis can alter the light element abundances through changes in the expansion history. In particular, increasing the expansion rate during Big Bang Nucleosynthesis causes weak reactions to freeze-out earlier, resulting in a higher helium-to-hydrogen ratio~\cite{Steigman:1977kc}. As these measurements conservatively exclude $\Delta N_{\rm Eff}^{\nu}>1$ at the 95\% confidence level~\cite{serpico}, we consider decays which occur later than $\tau \gsim 10^3$ to be the most attractive.


\bigskip

\section{Summary and Conclusions}

In this paper, we have proposed an alternative explanation for the lack of small scale power in the cosmic microwave background as reported by the South Pole Telescope (SPT), Atacama Cosmology Telescope (ACT), and Wilkinson Microwave Anisotropy Probe (WMAP). Instead of introducing an additional light neutrino species to account for these observations, we have considered the possibility that a small fraction ($\lsim 1\%$) of the universe's dark matter was produced through a non-thermal mechanism, such as the late-time decay of a much heavier state. As a consequence of their non-thermal origin, these dark matter particles possess significant kinetic energy, and thus contribute to the radiation density of the universe during the epoch prior to recombination. For appropriate choices of the decay time and masses, this scenario can impact the expansion history of the universe in a way that is indistinguishable from that predicted for an additional light neutrino species. We have considered constraints on this scenario from large scale structure and big bang nucleosynthesis, and in each case find acceptable regions of parameter space.

With the first cosmology results from Planck anticipated in early 2013, we will likely learn with relatively high precision the degree to which the small scale power of the cosmic microwave background is suppressed, and the number of effective neutrino species that would be required to produce this effect. If this number is found to be too large to be reconciled with the upper limits on the number of neutrino species present during big bang nucleosynthesis, it would help to further motivate late-time decay scenarios such as that presented in this paper.

\bigskip

{\it Acknowledgements}: We would like to thank Scott Dodelson for valuable discussions. DH is supported by the US Department of Energy and by NASA grant NAG5-10842. FSQ is supported by Coordena\c{c}\~{a}o de Aperfeicoamento de Pessoal de N\'{i}vel Superior (CAPES).

\end{document}